\documentclass[12pt]{article}
\usepackage{amssymb,amsmath,epsfig}
\allowdisplaybreaks

\begin{document}

\title{\bf Dynamical Instability of Shear-free Collapsing Star in Extended Teleparallel Gravity}
\author{Abdul Jawad \thanks{jawadab181@yahoo.com;~~abduljawad@ciitlahore.edu.pk}
and Shamaila Rani \thanks{shamailatoor.math@yahoo.com;~~drshamailarani@ciitlahore.edu.pk}\\
Department of Mathematics, COMSATS Institute of\\ Information
Technology, Lahore-54000, Pakistan.}

\date{}
\maketitle
\begin{abstract}
We study the spherically symmetric collapsing star in terms of
dynamical instability. We take the framework of extended
teleparallel gravity with non-diagonal tetrad, power-law form of
model presenting torsion and matter distribution as non-dissipative
anisotropic fluid. The vanishing shear scalar condition is adopted
to search the insights of collapsing star. We apply first order
linear perturbation scheme to metric, matter and $f(T)$ functions.
The dynamical equations are formulated under this perturbation
scheme to develop collapsing equation for finding dynamical
instability limits in two regimes such as Newtonian and
post-Newtonian. We obtain constraint free solution of perturbed time
dependent part with the help of vanishing shear scalar. The
adiabatic index exhibits the instability ranges through second
dynamical equation which depend on physical quantities such as
density, pressure components, perturbed parts of symmetry of star,
etc. We also develop some constraints on positivity of these
quantities and obtain instability ranges to satisfy the dynamical
instability condition.
\end{abstract}

\section{Introduction}

The gravitational collapse of self-gravitating objects has become
widely discussed phenomena in general relativity (GR) as well as in
modified theories of gravity. This contains the evolutionary
development and constancy of these objects during collapse process
and rests importantly at the center of structure formation. This
process occurs when a stable matter becomes unbalanced and
ultimately undergoes a collapse which results different structures
like stars, stellar groups and planets. In this way,
self-gravitating objects go across various dynamical states which
may be analyzed through dynamical equations. The dynamical
instability was firstly investigated by Chandrasekhar \cite{S17}
with the help of adiabatic index $\Gamma$ of a spherical star with
isotropic pressure. This index depicts the consequences of various
structural quantities of a fluid on the instability ranges. For
instability ranges in Newtonian and post-Newtonian regimes, Herrera
et al. \cite{S18} explored dissipative, non-adiabatic spherically
symmetric collapsing star.

The adiabatic index develops the instability ranges in GR as well as
in modified theories of gravity which induces that these ranges
depending on dark source terms in addition to usual terms. Under
different conditions for cylindrically and spherically symmetric
collapsing matters in $f(R)$ gravity, the instability ranges have
been explored through adiabatic index\cite{S19,FS19}. Taking
expansion-free condition, Skripkin \cite{S20} developed a model for
non-dissipative spherically symmetric fluid distribution with
isotropy and constant energy density and remarked that a Minkowskian
cavity is observed at a center of fluid. Under this condition, the
instability for spherically and cylindrically symmetric anisotropic
fluids in Newtonian, post-Newtonian regimes is explored in GR
\cite{S21} as well as $f(R)$ gravity \cite{S22}. In Brans-Dicke
gravity, Sharif and Manzoor \cite{rm} explored the instability
ranges of spherically symmetric collapsing star.

The physical aspects such as isotropy, radiation, anisotropy, shear,
dissipation, expansion are main sources of cause for the
gravitational evolution. Among these factors, the shear leads to the
formation of naked singularities. That is, it contribute to the
formation of an apparent horizon which results in a black hole of
the evolving cloud. Thus, the shear tensor occupies a direction of
well-motivation to study structure formation and its consequences on
the dynamical instability range of a self-gravitating body.

In context of extended teleparallel gravity (ETG) (or $f(T)$
gravity) which is the generalization of teleparallel gravity, the
gravitational collapse is discussed with and without expansion
scalar by Sharif and Rani \cite{sr,sr1}. They found that the
physical properties invade a vast impact of dynamical instability in
studying the self-gravitating objects with expansion. Without
expansion, they obtain the instability ranges for Newtonian (N) and
post-Newtonian (pN) regimes. In this paper, we assume shear-free
condition instead expansion-free and explore the instability ranges
of a collapsing star in ETG.

The scheme of the paper is given by: In section \textbf{2}, we give
the basics of ETG and provide the construction of field equations in
two ways, simple and covariant form. Section \textbf{3} contains the
basic equations for the static spherically symmetry. Also, junction
conditions are given for dynamical instability of a spherically
symmetric collapsing star in the context of ETG gravity. In the next
section, we represent perturbation scheme and ETG model and apply to
all matter, metric and $f(T)$ functions. In section \textbf{5}, we
formulate dynamical collapsing equation and found the instability
ranges in N and pN regimes. The last section summarizes the results
and elaborate the comparison.

\section{Extended Teleparallel Gravity}

In this section, we provide the basics such as tetrad field and the
Weitzenb\"{o}ck connection of ETG. We give field equations in simple
form as well as its covariant construction.

\subsection{Tetrad Field}

The geometry of ETG is unambiguously descried thorough an
orthonormal set having three spacelike and one timelike fields
called tetrad field. The trivial tetrad field has the form
$e_a=\delta_{a}^\mu\partial_\mu,~e^b=\delta^b_\mu dx^\mu$, where
$\delta^a_\mu$ named as the Kronecker delta. This is the simplest
field and less important due to zero torsion. The non-trivial tetrad
field allows non-zero torsion and grants the construction of
teleparallel as well as ETG theory. It is given by
\begin{equation}\label{1.1.6}
h_a={h_a}^\mu\partial_\mu,\quad h^b={h^b}_\nu dx^\nu
\end{equation}
satisfying the following properties
\begin{equation}\label{1.1.7}
{h^a}_\mu{h_b}^\mu=\delta^a_b,\quad
{h^a}_\mu{h_a}^\nu=\delta_\mu^\nu.
\end{equation}
The metric tensor is demonstrated as a by product of this field whih
is as follows
\begin{equation}\label{1.1.4}
g_{\mu\nu}=\eta_{ab}{h^a}_\mu{h^b}_\nu.
\end{equation}

\subsection{The Weitzenb\"{o}ck Connection}

The basic phenomenon of teleparallel gravity and ETG is the parallel
transport of tetrad field in Weitzenb\"{o}ck spacetime which is
carried out by the significant component Weitzenb\"{o}ck connection.

By applying covariant derivative w.r.t spacetime of tetrad field, we
get
\begin{equation}\label{1.2.8}
\Delta_{\nu}{h^a}_\mu=\partial_{\nu}{h^a}_\mu-{\widetilde{\Gamma}^\alpha}_{~\mu\nu}{h^a}_\alpha\equiv0,
\end{equation}
where
${\widetilde{\Gamma}^\alpha}_{~\mu\nu}={h_a}^{\alpha}\partial_\nu{h^a}_\mu$
is the Weitzenb\"{o}ck connection. We obtain torsion tensor by the
antisymmetric part of this connection as follows
\begin{equation}\label{1.2.10}
{T^\alpha}_{\mu\nu}={\widetilde{\Gamma}^\alpha}_{~\nu\mu}-{\widetilde{\Gamma}^\alpha}_{~\mu\nu}
={h_a}^{\alpha}(\partial_{\nu}{h^a}_{\mu}-\partial_{\mu}{h^a}_{\nu}),
\end{equation}
which is antisymmetric in its lower indices, i.e.,
${T^\alpha}_{\mu\nu}=-{T^\alpha}_{\nu\mu}$. This absolute
parallelism passed away rapidly the curvature of the Weitzenb\"{o}ck
connection identically. The following relation is also satisfied by
the Weitzenb\"{o}ck connection
\begin{equation}\label{1.2.12}
{\widetilde{\Gamma}^\alpha}_{~\mu\nu}={{\Gamma}^\alpha}_{\mu\nu}+
{K^\alpha}_{\mu\nu},
\end{equation}
here ${{\Gamma}^\alpha}_{\mu\nu},~{K^\alpha}_{\mu\nu}$ appear as the
usual Levi-Civita connection (torsionless) and the contorsion
tensor, respectively, which can be defined as follows
\begin{eqnarray}\label{1.2.13}
{{\Gamma}^\alpha}_{\mu\nu}&=&\frac{1}{2}g^{\alpha\rho}[g_{\rho\nu,\mu}+
g_{\rho\mu,\nu}-g_{\mu\nu,\rho}],\\\label{1.2.14}
{K^{\alpha}}_{\mu\nu}&=&\frac{1}{2}[{{T_\mu}^\alpha}_{\nu}+{{T_\nu}^\alpha}_{\mu}-
{T^\alpha}_{\mu\nu}].
\end{eqnarray}

\subsection{Field Equations}

To generalize the action of teleparallel gravity, we just substitute
a general function of torsion scalar by itself as follows
\cite{S3}-\cite{S4}
\begin{equation}\label{1.3.8}
\mathcal{S}=\frac{1}{2\kappa^2}\int h(f(T)+\mathcal{L}_m)d^{4}x.
\end{equation}
where, $h=\textmd{det}({h^a}_\lambda),~\mathcal{L}_m$ is the matter
Lagrangian and $f$ is the function of torsu=ion scalar. The torsion
scalar is defined as
\begin{equation}\label{1.3.5}
T={S_\alpha}^{\mu\nu}{T^\alpha}_{\mu\nu}.
\end{equation}
where
\begin{equation}\label{1.3.6}
{S_\alpha}^{\mu\nu}=\frac{1}{2}[-\frac{1}{2}(T^{\mu\nu}_{~~\alpha}
-T^{\nu\mu}_{~~\alpha}-T_{\alpha}^{~\mu\nu})+\delta^{\mu}_{\alpha}{T^{\theta\nu}}_{\theta}-
\delta^{\nu}_{\alpha}{T^{\theta\mu}}_{\theta}].
\end{equation}
called the superpotential tensor. Applying the variation of action
(\ref{1.3.8}) w.r.t tetrad field, we will get field equations as
follows
\begin{equation}\label{1.3.9}
[\frac{1}{h}\partial_{\mu}(h{h_a}^{\alpha}{S_\alpha}^{\mu\nu})
+{h_a}^{\alpha}{T^\lambda}_{\mu\alpha}{S_\lambda}^{\nu\mu}]f_{_T}
+{h_a}^{\alpha}{S_\alpha}^{\mu\nu}\partial_{\mu}T
f_{_{TT}}+\frac{1}{4}{h_a}^{\nu}f=\frac{1}{2}\kappa^{2}{h_a}^{\alpha}\mathcal{T}^{\nu}_{\alpha},
\end{equation}
where $f_{_T}$ is the first order derivative and $f_{_{TT}}$
represent second order derivative of $f$ with respect to $T$.

The field equations (\ref{1.3.9}) turn out to be extremely different
from Einstein's equations on account of tetrad components and
partial derivatives. Since tetrad are not completely eradicated
which causes difficulty to compare teleparallel gravity (and ETG)
with GR. To obtain equivalent description of field equations
(\ref{1.3.9}) with the other modified theories, we will apply
covariant formalism \cite{S31}. We replace all partial derivatives
in Eqs.(\ref{1.2.10}), (\ref{1.2.14}), (\ref{1.3.6}) by covariant
derivatives using the condition on metric tensor,
$\nabla_{\sigma}g_{\mu\nu}=0$, i.e., the compatibility of the metric
tensor, we get
\begin{eqnarray*}
&&{T^\alpha}_{\mu\nu}=h^{\alpha}_{a}
(\partial_{\mu}h^{a}_{\nu}-\Gamma^{\sigma}_{\mu\nu}h^{a}_{\sigma}-\partial_{\nu}h^{a}_{\mu}
+\Gamma^{\sigma}_{\mu\nu}h^{a}_{\sigma})=
h^{\alpha}_{a}(\nabla_{\mu}h^{a}_{\nu}-\nabla_{\nu}h^{a}_{\mu}),\\
&&{K^{\alpha}}_{\mu\nu}=h^{\alpha}_{a}\nabla_{\nu}h^{a}_{\mu},\quad
{S_\alpha}^{\nu\mu}=\eta^{ab}h^{\mu}_{a}\nabla_{\alpha}h^{\nu}_{b}+
\delta^{\nu}_{\alpha}\eta^{ab}h^{\sigma}_{a}\nabla_{\sigma}h^{\mu}_{b}
-\delta^{\mu}_{\alpha}\eta^{ab}h^{\sigma}_{a}\nabla_{\sigma}h^{\nu}_{b},
\end{eqnarray*}
where we have applied the following relations
\begin{eqnarray}\label{111a}
S^{\mu(\nu\alpha)}=T^{\mu(\nu\alpha)}=K^{(\mu\nu)\alpha}=0.
\end{eqnarray}
In this case, the Weitzenb\"{o}ck connection becomes zero, while the
Riemann tensor turns out to be
\begin{eqnarray}\nonumber
{R^\alpha}_{\mu\rho\nu}&=&\partial_{\rho}{\Gamma^\alpha}_{\mu\nu}-\partial_{\nu}{\Gamma^\alpha}_{\mu\rho}+
{\Gamma^\alpha}_{\sigma\rho}{\Gamma^\sigma}_{\mu\nu}-{\Gamma^\alpha}_{\sigma\nu}{\Gamma^\sigma}_{\mu\rho}\\\nonumber
&=&\nabla_{\nu}{K^\alpha}_{\mu\rho}-\nabla_{\rho}{K^\alpha}_{\mu\nu}+{K^\alpha}_{\sigma\nu}{K^\sigma}_{\mu\rho}
-{K^\alpha}_{\sigma\rho}{K^\sigma}_{\mu\nu}.
\end{eqnarray}
The corresponding Ricci tensor becomes
\begin{eqnarray*}
R_{\mu\nu}=\nabla_{\nu}{K^\alpha}_{\mu\alpha}-\nabla_{\alpha}{K^\alpha}_{\mu\nu}+
{K^\alpha}_{\rho\nu}{K^\rho}_{\mu\alpha}-{K^{\alpha}}_{\rho\alpha}{K^\rho}_{\mu\nu}.
\end{eqnarray*}
Using relations (\ref{111a}) along with
${S^\nu}_{\alpha\nu}=-2{T^\nu}_{\alpha\nu}=2{K^\nu}_{\alpha\nu}$, we
have
\begin{eqnarray}\label{116a}
R_{\mu\nu}&=&-\nabla^{\alpha}S_{\nu\alpha\mu}-g_{\mu\nu}\nabla^{\alpha}
{T^\rho}_{\alpha\rho}-{S^{\alpha\rho}}_{\mu}K_{\rho\alpha\nu},\\\label{gg}
R&=&-T-2\nabla^{\alpha}{T^\nu}_{\alpha\nu}.
\end{eqnarray}
The covariant derivative of torsion tensor in the last equation
shows the only difference between Ricci and torsion scalars.

After some calculations, we will get
\begin{equation}\label{1117a}
G_{\mu\nu}-\frac{1}{2}g_{\mu\nu}T=-\nabla^{\alpha}S_{\nu\alpha\mu}
-{S^{\sigma\alpha}}_{\mu}K_{\alpha\sigma\nu},
\end{equation}
where $G_{\mu\nu}=R_{\mu\nu}-\frac{1}{2}g_{\mu\nu}R$ appears as the
Einstein tensor. By using this in (\ref{1.3.9}), we attain the
required field equations in $f(T)$ gravity
\begin{equation}\label{118a}
f_{T}G_{\mu\nu}+\frac{1}{2}g_{\mu\nu}(f-Tf_T)+D_{\mu\nu}f_{TT}=\kappa^2\mathcal{T}_{\mu\nu},
\end{equation}
here $D_{\mu\nu}={S_{\nu\mu}}^{\alpha}\nabla_{\alpha}T$. It can be
observed that (\ref{118a}) has an equivalent structure such as
$f(R)$ gravity and reduces to GR for $f(T)=T$. Here, the trace of
the above equation is
\begin{equation}\label{119a}
Df_{TT}-(R+2T)f_T+2f=\kappa^2\mathcal{T},
\end{equation}
with $D={D^\nu}_{\nu}$ and $\mathcal{T}={\mathcal{T}^{\nu}}_{\nu}$.
The $f(T)$ field equations can also defined as
\begin{equation}\label{120a}
G_{\mu\nu}=\frac{\kappa^2}{f_T}(\mathcal{T}_{\mu\nu}^{m}+\mathcal{T}_{\mu\nu}^{T}).
\end{equation}
Here $\mathcal{T}_{\mu\nu}^{m}$ represents the matter fluid and
torsion contribution is
\begin{equation}\label{122a}
\mathcal{T}_{\mu\nu}^{T}=\frac{1}{\kappa^2}[-D_{\mu\nu}f_{TT}-
\frac{1}{4}g_{\mu\nu}(\mathcal{T}-Df_{TT}+Rf_{T})].
\end{equation}

\section{Basic Equations}

The general spherically symmetric metric in the interior region is
\begin{equation}\label{1c5}
ds^2_-=X^2dt^{2}-Y^2dr^{2}-R^2(d\theta^{2}+\sin^2\theta d\phi^{2}),
\end{equation}
where $X,~Y$ and $R$ are functions of $t$ and $r$. The line element
for exterior spacetime (the Schwarzschild metric) is \cite{S22}
\begin{equation}\label{21c5}
ds^2_+=\left(1-\frac{2M}{r}\right)d\upsilon^2+2drd\upsilon-r^2(d\theta^2+\sin^2\theta
d\phi^2),
\end{equation}
where $\upsilon$ is the retarded time and $M$ represents the total
mass of the bounded surface. Also, the anisotropic energy-momentum
tensor can be defined as follows
\begin{equation}\label{zc3}
\mathcal{T}^{\mu}_{~\nu}=(\rho+p_{\perp})u^{\mu}u_{\nu}-p_{\perp}\delta^{\mu}_{\nu}
+(p_r-p_{\perp})v^{\mu}v_{\nu},
\end{equation}
in the interior region while
$\rho=\rho(t,r),~p_r=p_r(t,r),~p_{\perp}=p_{\perp}(t,r)$. The four
velocity $u_\mu=\frac{1}{X}\delta^0_\mu$ and unit four vector
directed towards radial component $v_\mu=\frac{1}{Y}\delta^1_\mu$
satisfy the relations $u_\mu u^\mu=1,~v_\mu v^\mu=-1,~u_\mu
v^\mu=0$. We take the non-diagonal tetrad for the interior spacetime
as
\begin{center}${h^i}_{\mu}=\left(\begin{array}{cccc}
X & 0 & 0 & 0 \\
0 & Y\sin\theta \cos\phi & R\cos\theta\cos\phi & -R\sin\theta\sin\phi \\
0 & Y\sin\theta \sin\phi & R\cos\theta\sin\phi & R\sin\theta\cos\phi \\
0 & Y\cos\theta & -R\sin\theta & 0 \\
\end{array}\right).$
\end{center}

The description of gravitational collapsing star takes kinematics of
the dynamical equations of spherically symmetric models. This needs
acceleration, expansion, rotation and distortion or shear. Due to
shearfree condition, there have been many interesting results such
as this condition depicts the physical aspects of compact bodies in
the relativistic astrophysics phenomena. The shear scalar and tensor
are defined by
\begin{equation}\label{26c5++}
\sigma^2=\frac{1}{2}\sigma^{\alpha\beta}\sigma_{\alpha\beta},
\end{equation}
where
\begin{equation}\label{26c5+}
\sigma_{\alpha\beta}=V_{(\alpha;\beta)}+a_{(a}V_{\beta)}-\frac{1}{3}\Theta(g_{\alpha\beta}+V_\alpha
V_\beta), \quad \Theta=V^{\alpha}_{~;\alpha}, \quad
a=V_{(\alpha;\beta)}V^\beta.
\end{equation}
For interior spacetime, this tensor yields the following scalar
\begin{equation}\label{26c5}
\sigma=\frac{1}{X}\left(\frac{\dot{Y}}{Y}-\frac{\dot{R}}{R}\right).
\end{equation}
Using the tetrad along with Eq.(\ref{1c5}) in (\ref{120a}), the
field equations are
\begin{eqnarray}\nonumber
&&\left(\frac{2\dot{Y}}{Y}+\frac{\dot{R}}{R}\right)
\frac{\dot{R}}{R}-\left(\frac{X}{Y}\right)^2
\left[\frac{2R''}{R}+\left(\frac{R'}{R}\right)^2-\frac{2Y'R'}{YR}
-\left(\frac{Y}{R}\right)^2\right] \\\label{13c5}
&&=\frac{\kappa^2}{f_{_T}}\left[{\rho}X^{2}+
\frac{X^2}{\kappa^2}\left\{\frac{Tf_{_T}-f}{2}-\frac{1}{Y^2}\left(\frac{R'}{R}-
\frac{Y}{R}\right)f_T'\right\}\right],
\\\label{14c5}
&&2\left(\frac{\dot{R'}}{R}-\frac{\dot{R}X'}{RX}-\frac{\dot{Y}R'}{YR}\right)
=\frac{\dot{R}f_{_T}'}{R f_{_T}}, \\\label{142c5}
&&2\left(\frac{\dot{R'}}{R}-\frac{\dot{R}X'}{YR}-\frac{\dot{Y}R'}{YR}\right)
=\frac{\dot{T}}{T'}\left(\frac{R'}{R}-
\frac{Y}{R}\right)\frac{f_{_T}'}{f_{_T}},\\\nonumber
&&-\left(\frac{Y}{X}\right)^2\left[\frac{2\ddot{R}}{R}-\left(\frac{2\dot{X}}{X}
-\frac{\dot{R}}{R}\right) \frac{\dot{R}}{R}\right]
+\left(\frac{2X'}{X}+\frac{R'}{R}\right)\frac{R'}{R}-\left(\frac{Y}{R}\right)^2
\\\label{15c5}&&=\frac{\kappa^2}{f_{_T}}
\left[p_rY^{2}-\frac{Y^2}{\kappa^2}\left(\frac{Tf_{_T}-f}{2}
+\frac{\dot{R}\dot{T}f_{_T}'}{X^2RT'}\right)\right],\\\nonumber
&&-\left(\frac{R}{X}\right)^2\left[\frac{\ddot{Y}}{Y}+\frac{\ddot{R}}{R}-\frac{\dot{X}}{X}
\left(\frac{\dot{Y}}{Y}+\frac{\dot{R}}{R}\right)+\frac{\dot{Y}\dot{R}}{YR}\right]
+\left(\frac{R}{Y}\right)^2\left[\frac{X''}{X}+\frac{R''}{R}\right.\\\nonumber
&&-\left.\frac{X'Y'}{XY}+\left(\frac{X'}{X}-\frac{Y'}{Y}\right)\frac{R'}{R}\right]
=\frac{\kappa^2}{f_{_T}}\left[p_{\perp}R^2-\frac{R^2}{\kappa^2}\left\{\frac{Tf_{_T}-f}{2}+
\frac{1}{2}\right.\right.\\\label{16c5}
&&\times\left.\left.\left(\frac{1}{X^2}\left(\frac{\dot{Y}}{Y}+\frac{\dot{R}}{R}\right)\frac{\dot{T}}{T'}-
\frac{1}{Y^2}\left(\frac{X'}{X}+\frac{R'}{R}-\frac{Y}{R}\right)\right)f_{_T}'\right\}\right].
\end{eqnarray}
The torsion scalar takes the form
\begin{eqnarray}\nonumber
T&=&2\left[\frac{2}{XYR}\left(\frac{X'R'}{Y}-\frac{\dot{Y}\dot{R}}{X}\right)
-\frac{1}{X^2}\left(\frac{\dot{R}}{R}\right)^2+\frac{1}{Y^2}\left(\frac{R'}{R}\right)^2+
\frac{1}{R^2}\right.\\\label{17c5}&-&\left.\frac{2}{YR}\left(\frac{X'}{X}+\frac{R'}{R}\right)\right].
\end{eqnarray}
Also, from Eqs.(\ref{14c5}) and (\ref{142c5}), we obtain a
relationship as follows
\begin{eqnarray}\label{xyc5}
\frac{\dot{R}}{R}=\frac{\dot{T}}{T'}\left(\frac{R'}{R}+\frac{Y}{R}\right).
\end{eqnarray}
To analyze the properties of collapsing star, the non-trivial
contracted identities yield the dynamical equations which are very
useful. These are given by
\begin{eqnarray}\label{18c5}
\left(\overset{m}{\mathcal{T}^{\mu\nu}}+\overset{T}{\mathcal{T}^{\mu\nu}}\right)_{;\nu}u_{\mu}=0,\quad
\left(\overset{m}{\mathcal{T}^{\mu\nu}}+\overset{T}{\mathcal{T}^{\mu\nu}}\right)_{;\nu}
v_{\mu}=0.
\end{eqnarray}
Using these equation, the dynamical equations become
\begin{eqnarray}\nonumber
&&\left[\frac{\dot{\rho}}{X}+(\rho+p_r)\frac{\dot{Y}}{XY}+2(\rho+p_{\perp})
\frac{\dot{R}}{XR}\right]+\frac{X}{\kappa^2}\left[\left(\frac{Tf_{_T}-f}{2X^2}-
\frac{\dot{R}T'f_{_T}'}{X^2Y^2R\dot{T}}\right)_{,0}\right.\\\nonumber&&
+ \left.\frac{\dot{X}}{X^3}(Tf_{_T}-f)+\left(\frac{\dot{R}f_{_T}'}
{X^2Y^2R}\right)_{,1}-
\frac{1}{X^2R}\left\{\frac{2\dot{R}}{XY^2}\left(\frac{\dot{X}T'}{\dot{T}}-2X'\right)
+\frac{\dot{R}}{X^2}\right.\right.\\\label{19c5}&&\times\left.\left.
\left(\frac{2\dot{Y}}{Y}+\frac{\dot{R}}{R}\right)\frac{\dot{T}}{T'}+\frac{\dot{R}}{Y^2}
\left(\left(\frac{{\dot{R}}}{R}+\frac{\dot{Y}}{Y}\right)
\frac{T'}{\dot{T}}-\frac{Y'}{Y}-\frac{2R'}{R}\right)\right\}f_{_T}'\right]=0,
\\\nonumber
&&\left[\frac{p_r'}{Y}+(\rho+p_r)\frac{X'}{XY}+2(p_r-p_{\perp})\frac{R'}{YR}\right]+\frac{Y}{\kappa^2}
\left[\left(\frac{\dot{R}f_{_T}'}{X^2Y^2R}\right)_{,0}-\frac{1}{Y^2}\right.\\\nonumber
&&\times\left.\left(\frac{Y'}{Y}
+\frac{R'}{R}\right)(Tf_{_T}-f)+\left(\frac{f-Tf_{_T}}{2Y^2}-
\frac{\dot{R}\dot{T}f_{_T}'}{X^2Y^2RT'}\right)_{,1}+
\frac{1}{XY^2R}\right.\\\nonumber&&\left.\times\left\{-\frac{X'\dot{R}T'}{Y^2\dot{T}}+
\frac{\dot{R}}{X}\left(\frac{3\dot{Y}}{Y}+\frac{\dot{X}}{X}\right)
+2\frac{\dot{R}^2}{XR}
\left(1-\frac{1}{Y^2}\right)-\frac{\dot{R}}{XY}\right.\right.
\\\label{20c5}&&\left.\left.\times\left(Y'+\frac{Y'}{Y^2}+
\frac{X'}{XY}\right)\frac{\dot{T}}{T'}\right\}f_T'\right]=0.
\end{eqnarray}

\subsection{Junction Conditions}

In order to join smoothly the interior and exterior spacetimes over
a hypersurface $\Sigma^{(e)}$, we apply junction conditions. The
collapse problems are dealt by Darmois junction conditions in
appropriate manner. These conditions require the continuity of
intrinsic and extrinsic curvatures over the hypersurface, i.e.,
$(ds^2)_{\Sigma}=(ds^2_{-})_{\Sigma}=(ds^2_{+})_{\Sigma}$ and
$\mathcal{K}_{ab}=\mathcal{K}_{ab}^{-}=\mathcal{K}_{ab}^{+}$,
respectively. The Misner-Sharp mass function is given by
\begin{equation*}
m(t,r)=\frac{R}{2}(1+g^{\mu\nu}R_{,\mu}R_{,\nu}),
\end{equation*}
where a spherical object of radius $R$ contributes total energy and
contributes to study Darmois junction conditions. For
Eq.(\ref{1c5}), it takes the form
\begin{equation}\label{22c5}
m(t,r)=\frac{R}{2}\left(1+\frac{\dot{R}^2}{X^2}
-\frac{R'^2}{Y^2}\right).
\end{equation}

In order to match exterior region with interior, it requires that
$r=r_{\Sigma^{(e)}}=constant$ on the boundary surface $\Sigma^{(e)}$
\cite{S22,18c5} which results
\begin{eqnarray}\label{23c5}
M&\overset{\Sigma^{(e)}}{=}&m(t,r),
\end{eqnarray}
\begin{eqnarray}\nonumber
2\left(\frac{\dot{R'}}{R}-\frac{\dot{R}X'}{RX}-\frac{\dot{Y}R'}{YR}\right)
&\overset{\Sigma^{(e)}}{=}&
-\frac{Y}{X}\left[\frac{2\ddot{R}}{R}-\left(\frac{2\dot{X}}{X}-
\frac{\dot{R}}{R}\right) \frac{\dot{R}}{R}\right]\\\label{24c5}
&+&\frac{X}{Y}\left[\left(\frac{2X'}{X}+\frac{R'}{R}\right)\frac{R'}{R}
-\left(\frac{Y}{R}\right)^2\right].
\end{eqnarray}
Substituting the field equations (\ref{14c5}) and (\ref{15c5}) in
the above equation, we get
\begin{equation}\label{25c5}
-p_r\overset{\Sigma^{(e)}}{=}\frac{\mathcal{T}^{T}_{11}}{Y^2}
-\frac{\mathcal{T}^{T}_{01}}{XY}=\frac{f(T_c)}{2},
\end{equation}
where $f(T_c)$ represents a constant value for constant torsion
scalar $T_c$.

\section{Linear Perturbation Strategy and Power-law $f(T)$ Model}

In $f(R)$ gravity, the gravitational collapse is widely discussed
taking power-law form of model which is simply generalizes GR. We
take particular $f(T)$ model in power-law form analogy to $f(R)$
model \cite{S22} like $f(R)=R+\gamma R^2$ to analyze the evolution
of collapsing star. The power-law $f(T)$ model has contributed as a
most viable model due to its simple form and we may directly compare
our results with GR. We assume the ETG model as follows
\begin{equation}\label{29c5}
f(T)=T+\delta T^2,
\end{equation}
where $\delta$ is an arbitrary constant. For this model, we obtain
accelerated expansion universe in phantom phase, possibility of
realistic wormhole solutions and instability conditions for a
collapsing star. We assume the linear perturbation strategy to
construct the dynamical equations in order to explore instability
ranges for the underlying scenario. For this purpose, we assume the
system initially in static equilibrium. That is, metric and matter
parts are at zero order perturbation are only radial dependent only
which also become time dependent for the first order perturbations.
These perturbations are described as follows \cite{S19}-\cite{sr1},
\cite{18c5}-\cite{5c5}
\begin{eqnarray}\label{41cc5}
X(t,r)&=&X_0(r)+\epsilon \Pi(t)x(r),\\\label{42cc5}
Y(t,r)&=&Y_0(r)+\epsilon \Pi(t)y(r),\\\label{43cc5}
R(t,r)&=&R_0(r)+\epsilon \Pi(t)c(r),\\\label{44cc5}
\rho(t,r)&=&\rho_0(r)+\epsilon {\hat{\rho}}(t,r),\\\label{45cc5}
p_r(t,r)&=&p_{r0}(r)+\epsilon {\hat{p}_r}(t,r), \\\label{46cc5}
p_{\perp}(t,r)&=&p_{\perp0}(r)+\epsilon{\hat{p}_\perp}(t,r),\\\label{48cc5}
m(t,r)&=&m_0(r)+\epsilon \hat{m}(t,r), \\\label{49'cc5}
T(t,r)&=&T_0(r)+\epsilon \Pi(t)e(r),
\end{eqnarray}
where the quantities with zero subscript denotes static parts of
corresponding functions and $0<\epsilon\ll1$. The perturbed
vanishing shear scalar and $f(T)$ model takes the form
\begin{eqnarray}\label{a+}
f(T)&=&T_0(1+\delta T_0)+\epsilon \Pi e(1+2\delta
T_0),\\\label{51'cc5} f_{_T}(T)&=&1+2\delta T_0+2\epsilon \delta \Pi
e,\\\label{a} \frac{y}{Y_0}&=&\frac{c}{R_0}.
\end{eqnarray}

Taking into account shear-free condition ($\sigma=0$),
Eq.(\ref{26c5}) yields $\frac{\dot{Y}}{Y}=\frac{\dot{R}}{R}$. The
solution of this equation turns out as $Y=\alpha R$ where $\alpha$
is an arbitrary function of $r$ taken as 1 without loss of
generality and using the freedom to rescale the radial coordinate,
we take $R_0=r$ which is also the Schwarzschild coordinate. The
condition under which an initially shear-free flow remains
shear-free all along the evolution, has been studied by Herrera et
al. \cite{Her}. One of the consequences of such a study is that the
pressure anisotropy may affect the propagation in time, of the
shear-free condition. The shear-free condition is unstable, in
particular, against the presence of pressure anisotropy. An
expansion scalar and a scalar function insured the departures from
the shear–free condition for the geodesic case. These scalars are
defined in purely physical variables such as in terms of the Weyl
tensor, anisotropy of pressure and the shear viscosity. It is
remarked that one can consider such a case that pressure anisotropy
and density inhomogeneity are present in a way that the scalar
function appearing in orthogonal splitting of Reimann tensor
vanishes, implying non-homogeneous anisotropic stable shear-free
flow. Since we are dealing with fluid evolving under shearfree
condition, so we shall make use of this condition while evaluating
the components of field equations and also in conservation
equations.

Now we evaluate zero order as well as first order configurations.
The zero order perturbation of the field equations
(\ref{13c5})-(\ref{16c5}) is given by
\begin{eqnarray}\label{50cc5}
&&\frac{1}{1+2\delta
T_0}\left[\kappa^2\rho_0+\delta\left(\frac{T_0^2}{2}-\frac{2cT_0'^2}{erY_0^2}\right)\right]
=\frac{1}{(Y_0r)^2}\left(2r\frac{Y_0'}{Y_0}+Y_0^2-1\right),
\\\label{51cc5} &&\frac{1}{1+2\delta T_0}\left[\kappa^2
p_{r0}-\delta\frac{T_0^2}{2}\right]=\frac{1}{(Y_0r)^2}\left(2r\frac{X_0'}{X_0}-Y_0^2+1\right),\\\nonumber
&&\frac{1}{1+2\delta T_0}\left[\kappa^2
p_{\perp0}-\delta\frac{T_0^2}{2}+\frac{\delta
T_0'}{Y_0^2}\left(\frac{X_0'}{X_0}+\frac{cT_0'}{re}\right)\right]=\frac{1}{Y_0^2}\left[\frac{X_0''}{X_0}
-\frac{X_0'}{X_0}\frac{Y_0'}{Y_0}\right.\\\label{52cc5}
&&\left.+\frac{1}{r}
\left(\frac{X_0'}{X_0}-\frac{Y_0'}{Y_0}\right)\right].
\end{eqnarray}
The first dynamical equation (\ref{19c5}) fulfills the zero order
perturbation identically while second dynamical equation
(\ref{20c5}) becomes
\begin{equation}\label{j3c5}
\frac{2}{r}(p_{r0}-p_{\perp0})+p_{r0}'+\frac{X_0'}{X_0}(\rho_0+p_{r0})
-\frac{\delta}{\kappa^2}\left[T_0T_0'+\frac{T_0^2}{r}+\frac{2cX_0'T_0'^2}{eX_0Y_0^2r}\right]=0.
\end{equation}
In static background, the matching condition, mass function and
torsion scalar contribute for static equilibrium as
\begin{eqnarray}\label{63'ccc5}
p_{r0}\overset{\Delta^{(e)}}{=}\frac{\delta T_0^2}{2\kappa^2}, \quad
m_0(r)=\frac{r}{2}\left(1-\frac{1}{Y^2_0}\right),\\\label{63'cc5}
T_{0}=\frac{2}{r}\left[\frac{1}{r}\left(1-\frac{3}{Y_0}\right)
-\frac{1}{Y_0}\left(\frac{2X_0'}{X_0}+\frac{1}{r}\right)\left(1-\frac{1}{Y_0}\right)\right].
\end{eqnarray}

Applying the perturbed quantities to the field equations, we obtain
\begin{eqnarray}\nonumber
&-&\frac{2\Pi}{Y_0^2}\left[\left(\frac{c}{r}\right)''-\frac{1}{r}
\left(\frac{y}{Y_0}\right)'
-\left(\frac{Y_0'}{Y_0}-\frac{3}{r}\right)\left(\frac{c}{r}\right)'
\right]\\\nonumber &=&\frac{2\Pi
c}{Y_0^2r^3}\left(2r\frac{Y_0'}{Y_0}+Y_0^2-1\right)+\frac{\kappa^2
\hat{\rho}}{1+2\delta T_0}+\frac{\Pi\delta}{(1+2\delta
T_0)}\left[T_0e-\frac{2\kappa^2 \rho_0 y}{1+2\delta
T_0}\right.\\\label{53cc5} &-&\left.\frac{\delta T_0^2y}{1+2\delta
T_0}-\frac{cT_0'}{eY_0^2r}\left(2e'-\frac{4\delta T_0'e}{1+2\delta
T_0}-\frac{4cT_0'}{r}\right)\right],\\\label{54cc5}
&&\left(\frac{c}{r}\right)'-\left(\frac{X_0'}{X_0}-\frac{1}{r}\right)\frac{c}{r}-\frac{c}{r^2}=\frac{\delta
T_0'c}{r(1+2\delta T_0)},
\\\label{54ccc5}
&&\left(\frac{c}{r}\right)'-\left(\frac{X_0'}{X_0}-\frac{1}{r}\right)\frac{c}{r}-\frac{c}{r^2}=
\frac{e(1-Y_0)\delta}{r(1+2\delta T_0)},
\\\nonumber
&-&\frac{2\ddot{\Pi}Y_0^2c}{rX_0^2}+
\frac{2\Pi}{r}\left[\left(\frac{x}{X_0}\right)'+
\left(r\frac{X_0'}{X_0}+1\right)\left(\frac{c}{r}\right)'\right]\\\nonumber
&=&\frac{\kappa^2 Y_0^2\hat{p_r}}{1+2\delta T_0}+\frac{2\kappa^2\Pi
Y_0^2}{1+2\delta T_0}\left(\frac{c}{r}-\frac{\delta e}{1+2\delta
T_0}\right)p_{r0}-\frac{\delta \Pi T_0Y_0^2}{1+2\delta T_0}
\left(e+\frac{cT_0}{r}\right. \\\label{55cc5} &-&\left.\frac{\delta
Y_0T_0e}{1+2\delta T_0}\right), \\\nonumber
&-&\frac{2c\ddot{\Pi}}{rX_0^2}+\frac{\Pi}{Y_0^2}
\left[\left(\frac{x}{X_0}\right)''
+\left(\frac{c}{r}\right)''+\left(\frac{2X_0'}{X_0}-\frac{Y_0'}{Y_0}+\frac{1}{r}\right)
\left(\frac{x}{X_0}\right)'\right.\\\nonumber &-&\left.
\left(\frac{Y_0'}{Y_0}-\frac{1}{r}\right)
\left(\frac{c}{r}\right)'\right]=\frac{\kappa^2\hat{p}_{\perp}}{1+2\delta
T_0} \\\nonumber&-&\frac{2\Pi e\delta\kappa^2}{(1+2\delta
T_0)^2}p_{\perp0}+\frac{2\Pi c}{rY_0^2}\left[\frac{X_0''}{X_0}
-\frac{X_0'}{X_0}\frac{Y_0'}{Y_0}+\frac{1}{r}
\left(\frac{X_0'}{X_0}-\frac{Y_0'}{Y_0}\right)\right]\\\nonumber
&+&\frac{\delta \Pi}{1+2\delta T_0}\left[\frac{e\delta
T_0^2}{1+2\delta T_0}-eT_0 +\frac{X_0'e'}{X_0Y
_0^2}-\frac{2cT_0'X_0'}{rX_0Y_0^2}-\frac{2eT_0'X_0'\delta}{X_0Y_0^2(1+2\delta
T_0)}\right. \\\label{12''c5}
&-&\left.\frac{aX_0'T_0'}{X_0^2Y_0^2}+\frac{x'T_0'}{X_0Y_0^2}+\frac{e'c^2T_0'}{er^2Y_0^2}-\frac{2cT_0'}{r^2Y_0^2}
-\frac{2c\delta T_0'^2}{rY_0^2(1+2\delta T_0)}\right].
\end{eqnarray}
Applying Eqs.(\ref{41cc5})-(\ref{46cc5}) to the Bianchi identities,
it follows that
\begin{eqnarray}\label{j4c5}
\dot{\hat{\rho}}+\left[(3\rho_0+p_{r0}+2p_{\perp0})\frac{c}{r}+J_0\right]\dot{\Pi}=0,
\end{eqnarray}
where
\begin{eqnarray}\nonumber
J_0&=&\frac{1}{\kappa^2}\left[\delta
T_0e-\frac{4cT_0'\delta}{erY_0^2}\left(e'-T_0'\left(\frac{x}{X_0}
+\frac{2c}{r}\right)\right)+\frac{2\delta}{X_0^2}\left(\frac{cT_0'}{X_0^2Y_0^2r}\right)_{,1}
\right.\\\nonumber&-&\left.\frac{4c\delta
T_0'}{X_0Y_0^2r}\left(\frac{xT_0'}{e}-2X_0'\right)+\frac{2c\delta
T_0'}{rY_0^2}\left\{\frac{2}{r}+\frac{Y_0'}{Y_0}-\frac{2cT_o'}{er}\right\}\right].
\end{eqnarray}
Integrating Eq.(\ref{j4c5}) with respect to time, we have
\begin{eqnarray}\label{62cc5}
\hat{\rho}&=&-\left[(3\rho_0+p_{r0}+2p_{\perp0})
\frac{c}{r}+J_0\right]\Pi.
\end{eqnarray}
The perturbed second Bianchi identity is given by
\begin{eqnarray}\nonumber
&&p_{r0}'-(\rho_0+p_{r0})\left(\frac{x}{X_0}\right)'\frac{r}{c}+(\rho_0+p_{r0})\frac{X_0'}{X_0}
-(\hat{\rho}+\hat{p}_r)\frac{X_0'r}{X_0\Pi r}+J_1\\\label{j9c5}
&&-2(p_{r0}-p_{\perp0})\left(\frac{r}{c}\left(\frac{c}{r}\right)'
-\frac{1}{r}\right)-(\hat{p}_r-\hat{p}_{\perp})\frac{2}{\Pi
c}-\frac{c}{\Pi r}\hat{p}_r'=0,
\end{eqnarray}
where
\begin{eqnarray}\nonumber
J_1&=&\frac{r\delta}{c\kappa^2}\left[Y_0^2\left(\frac{T_0
e}{Y_0^2}-\frac{T_0^2c}{rY_0^2}\right)_{,1}-\frac{2
T_0'c}{rX_0^2}\frac{\ddot{\Pi}}{\Pi}+\frac{2T_0Y_0'e}{Y_0}-
\frac{3cY_0'T_0^2}{rY_0}\right.\\\nonumber
&+&\left.\left(\frac{cY_0}{r}\right)'\frac{T_0^2}{Y_0}-\frac{2cT_0^2}{r^2}
+\frac{2T_0e}{r}+T_0^2\left(\frac{c}{r}\right)'+\frac{2cx'T_0'^2}{erX_0Y_0^2}\right.
\\\nonumber&+&
\left.+\frac{2cX_0'T_0'}{erX_0Y_0^2}\left(1-\frac{m_0}{r}\right)'\left(1-\frac{m_0}{r}\right)
\left(2e'-\frac{xT_0'}{X_0}-\frac{4cT_0'}{r}\right)\right].
\end{eqnarray}

The junction condition, torsion scalar and mass function are
\begin{eqnarray}\label{622c5}
\hat{p_r}&\overset{\Sigma^{(e)}}{=}&\frac{\delta}{\kappa^2}\left(\Pi
T_0e-\frac{2T_0'\dot{\Pi}c}{rX_0Y _0}\right),\\\nonumber
e&=&\frac{4}{rY_0^2}\left[\frac{1}{X_0Y_0}\left\{x'+X_0'c'-\frac{X_0'}{Y_0}\left(y+
\frac{x}{X_0}+\frac{y}{Y_0}+\frac{c}{r}\right)\right\}\right.
\\\label{6222c5}&-&\left.
\frac{cY_0}{r^2}+\frac{xX_0'}{X_0^2}-\frac{x'}{X_0}-\left(\frac{c}{r}\right)'
+\left(\frac{y}{Y_0}+\frac{c}{r}\right)\left(\frac{X_0'}{X_0}+\frac{2}{r}\right)\right],\\\label{226c5}
\bar{m}&=&-\frac{\Pi}{Y_0^2}\left[r\left(c'-\frac{y}{Y_0}\right)+\frac{c}{2}(1-Y
_0^2) \right].
\end{eqnarray}
Substituting zero order and first order matching conditions in
Eq.(\ref{55cc5}), we have
\begin{equation}\label{j5c5}
\ddot{\Pi}\overset{\Sigma^{(e)}}{=}0,
\end{equation}
yields the general solution of this equation is
\begin{equation}\label{ac5}
\Pi(t)=h_1t+h_2,
\end{equation}
where $h_1$ and $h_2$ are arbitrary constants. It is remarked that
we do not need to impose any extra condition on this solution due to
vanishing shear scalar condition for collapse to occur. In
\cite{S19,FS19, S22}-\cite{sr1}, we need to apply some constraint
for the static solution in order to discuss instability analysis.

\section{Collapse Equation and Dynamical Instability}

In this section, we construct the collapse equation in order to work
for dynamical instability in different regimes with shear-free
condition. The Harison-Wheeler equation of state is used in this
regard which is given by \cite{ac5}
\begin{equation}\label{j7c5}
{\hat{p}_{ir}}=\hat{\rho} \frac{p_{i0}}{\rho_0+p_{i0}}\Gamma.
\end{equation}
We use this index in order to examine instability ranges in the
context of ETG with $\sigma=0$. The adiabatic index $\Gamma$ finds
the rigidity of the fluid and evaluates the change of pressure to
corresponding density. Substituting the value of $\hat{\rho}$, it
follows that
\begin{eqnarray}\label{j8c5}
{\hat{p}_r}&=&-\Pi \left[\frac{2c}{r}\frac{\rho_0+p_{\perp0}}{\rho_0
+p_{r0}}p_{r0}+\frac{c}{r}p_{r0}+\frac{p_{r0}}{\rho_0+p_{r0}}J_0\right]\Gamma,\\\label{j88c5}
{\hat{p}_\perp}&=&-\Pi \left[\frac{c}{r}\frac{\rho_0+p_{r0}}{\rho_0
+p_{\perp0}}p_{\perp0}+\frac{2c}{r}p_{\perp0}+\frac{p_{\perp0}}{\rho_0+p_{\perp0}}J_0\right]\Gamma.
\end{eqnarray}
Finally, the collapse equation used to analyze instability ranges of
collapsing star in N and pN regimes, we insert all the corresponding
values in Eq.(\ref{j9c5}), we have
\begin{eqnarray}\nonumber
&&\frac{\delta T_0
T'_0}{\kappa^2}-(\rho_0+p_{r0})\left[\frac{xr}{cX_0}-\frac{X'_0}{X_0}\right]+
(\Gamma+1)\left(\frac{X'_0r}{X_0c}\right)\left[\frac{2c}{r}(\rho_0+p_{\perp0})+J_0\right.
\\\nonumber&&\left.+\frac{c}{r}(\rho_0+p_{r0})\right]
+J_1-2(p_{r0}-p_{\perp0})\left(\frac{r}{c}\left(\frac{c}{r}\right)'-\frac{1}{r}\right)
+\Gamma\left[\frac{c}{r}\left(p_{r0}\right.\right.\\\nonumber&&\left.\left.-\frac{\rho_0+p_{r0}}{\rho_0
+p_{\perp0}}p_{\perp0}\right)+\frac{2c}{r}\left(\frac{\rho_0+p_{\perp0}}{\rho_0
+p_{r0}}p_{r0}-p_{\perp0}\right)+\left(\frac{p_{r0}}{\rho_0+p_{r0}}-
\frac{p_{\perp0}}{\rho_0+p_{\perp0}}\right)\right.
\\\label{j12c5}&&\times\left. J_0\right]+\frac{\Gamma
r}{c}\left[\frac{c}{r}p_{r0}+\frac{2c}{r}\frac{\rho_0+p_{\perp0}}{\rho_0
+p_{r0}}p_{r0}+\frac{p_{r0}}{\rho_0+p_{r0}}J_0\right]_{,1}=0.
\end{eqnarray}
We consider the N and pN approximations to obtain the instability
ranges in the following.

\subsection*{Newtonian limit}

The N approximations are given by
\begin{equation}\nonumber
X_0=1=Y_0, \quad \rho_0\gg p_{r0},\quad \rho_0\gg p_{\perp0},\quad
p_{r0}\gg p_{\perp0}.
\end{equation}
Inserting these conditions as well as Eq.(\ref{ac5}) in
(\ref{j12c5}), the collapse equation yields
\begin{eqnarray}\label{j133c5}
\frac{\delta T_0 T'_0}{\kappa^2}-\frac{xr
\rho_0}{c}+J_{1(N)}-2p_{r0}\left(\frac{r}{c}\left(\frac{c}{r}\right)'-\frac{1}{r}\right)+
\Gamma\left[\frac{3cp_{r0}}{r}+\frac{3r}{c}\left(\frac{cp_{r0}}{r}\right)_{,1}\right]=0,
\end{eqnarray}
where
\begin{eqnarray}\nonumber
J_{1(N)}=\frac{r\delta}{c\kappa^2}\left[\left(eT_0-\frac{cT_0^2}{r}\right)_{,1}+
2\left(\frac{c}{r}\right)'T_0^2-\frac{2cT_0^2}{r^2}+\frac{2T_0e}{r}
+\frac{2cx'T_0'^2}{er}\right].
\end{eqnarray}
To check the instability range for N approximation, we obtain
\begin{eqnarray}\label{j13c5}
\Gamma<\frac{1}{I_0}\left[-\frac{\delta T_0 T'_0}{\kappa^2}+\frac{xr
\rho_0}{c}-J_{1(N)}+2p_{r0}\left(\frac{r}{c}\left(\frac{c}{r}\right)'-\frac{1}{r}\right)\right],
\end{eqnarray}
where
$I_0=\frac{3cp_{r0}}{r}+\frac{3r}{c}\left(\frac{cp_{r0}}{r}\right)_{,1}$.
This equation expresses the dependence of adiabatic index on torsion
terms along with physical properties such as anisotropic pressure
and energy density. As long as the above inequality maintains, the
collapsing system stays unstable. To be hold for dynamical
instability condition the terms in collapse equation should be
positive in the inequality. It requires that
\begin{eqnarray}\nonumber
-\frac{\delta T_0 T'_0}{\kappa^2}+\frac{xr
\rho_0}{c}-J_{1(N)}+2p_{r0}\left(\frac{r}{c}\left(\frac{c}{r}\right)'-\frac{1}{r}\right)>0,
\end{eqnarray}
where $I_0>0$ holds. In GR \cite{S17}, the adiabatic index
represents a numerical value corresponding to dynamical instability
of an isotropic sphere. This is given by
\begin{itemize}
\item It is found that the dynamical stability is achieved for
$\Gamma>\frac{4}{3}$ when the weight of outer layers is weaker in
contrast to the pressure in star.
\item The case $\Gamma=\frac{4}{3}$ corresponds to the hydrostatic equilibrium condition.
\item When the weight of outer layers increases very fast as compared to pressure inside the star gives the
collapse and $\Gamma<\frac{4}{3}$ constitutes the dynamical
instability.
\end{itemize}
The expressions in Eq.(\ref{j13c5}) constitute the following
possibilities
\begin{eqnarray}\nonumber
(i) \quad\left[-\frac{\delta T_0 T'_0}{\kappa^2}+\frac{xr
\rho_0}{c}-J_{1(N)}+2p_{r0}\left(\frac{r}{c}\left(\frac{c}{r}\right)'-\frac{1}{r}\right)\right]=I_0,\\\nonumber
(ii) \quad\left[-\frac{\delta T_0 T'_0}{\kappa^2}+\frac{xr
\rho_0}{c}-J_{1(N)}+2p_{r0}\left(\frac{r}{c}\left(\frac{c}{r}\right)'-\frac{1}{r}\right)\right]<I_0,\\\nonumber
(iii) \quad\left[-\frac{\delta T_0 T'_0}{\kappa^2}+\frac{xr
\rho_0}{c}-J_{1(N)}+2p_{r0}\left(\frac{r}{c}\left(\frac{c}{r}\right)'-\frac{1}{r}\right)\right]>I_0.
\end{eqnarray}
In the first possibility together with Eq.(\ref{j13c5}), we obtain
the instability range as $0<\Gamma<1$. The case $(ii)$ yields that
faction in Eq.(\ref{j13c5}) always less than 1 but depending on the
values of dynamical terms. The third case $(iii)$ represents the
different numerical values for different values of dynamical terms.
Thus, it shows that the collapsing star remains unstable for
$\Gamma>1$. Also, Eq.(\ref{j133c5}) depicts that the adiabatic index
contains the physical quantities like $f(R)$ gravity
\cite{FS19,S22}.

\subsection*{Post-Newtonian limit}

We consider the following pN approximations in terms of metric
component expressions given by
\begin{eqnarray}\label{j14c5}
X_0=1-\frac{m_0}{r},\quad Y_0=1+\frac{m_0}{r},
\end{eqnarray}
upto order $\frac{m_0}{r}$. Introducing these approximations in
Eq.(\ref{j12c5}), we obtain
\begin{eqnarray}\nonumber
&&\frac{\delta T_0
T'_0}{\kappa^2}-\left[\frac{xr}{c}\left(1+\frac{m_0}{r}\right)-\frac{m_0}{r^2}\right]
(\rho_0+p_{r0})+J_{1(pN)}-2(p_{r0}-p_{\perp0})\\\nonumber&&\times\left[\frac{r}{c}
\left(\frac{c}{r}\right)'-\frac{1}{r}\right]+
\frac{(\Gamma+1)m_0}{rc}\left[\frac{c}{r}(3\rho_0+p_{r0}+2p_{\perp0})+J_{0(pN)}\right]
\\\nonumber&&+\Gamma\left[\frac{c}{r}\left(p_{r0}-\frac{\rho_0+p_{r0}}{\rho_0
+p_{\perp0}}p_{\perp0}\right)+\frac{2c}{r}\left(\frac{\rho_0+p_{\perp0}}{\rho_0
+p_{r0}}p_{r0}-p_{\perp0}\right)\right.
\\\nonumber&&\left.
+\left(\frac{p_{r0}}{\rho_0+p_{r0}}-
\frac{p_{\perp0}}{\rho_0+p_{\perp0}}\right)
J_{0(pN)}\right]+\frac{\Gamma r}{c} \left[\frac{c}{r}p_{r0}
+\frac{2c}{r}\frac{p_{\perp0}+\rho_0}{\rho_0
+p_{r0}}p_{r0}\right.\\\label{j155c5}&&\left.+J_{0(pN)}\frac{p_{r0}}{\rho_0+p_{r0}}\right]_{,1}=0,
\end{eqnarray}
where $J_{1(pN)}$ and $J_{0(pN)}$ are those terms which belong to pN
regime in $J_1$ and $J_0$ expressions are given below.
\begin{eqnarray}\nonumber
J_{0(pN)}&=&\frac{1}{\kappa^2}\left[\delta
T_0e-\frac{4cT_0'\delta}{er}\left(1-\frac{2m_0}{r}\right)\left\{e'-T_0'\left(x\left(1+\frac{m_0}{r}\right)
+\frac{2c}{r}\right)\right\}\right.\\\nonumber&+&\left.2\delta\left(1+\frac{2m_0}{r}\right)\left(\frac{cT_0'}{r}\right)_{,1}
-\frac{4c\delta
T_0'}{r}\left(1-\frac{m_0}{r}\right)\left(\frac{xT_0'}{e}+\frac{2m_0}{r^2}\right)\right.\\\nonumber&+&\left.\frac{2c\delta
T_0'}{r}\left(1-\frac{2m_0}{r}\right)\left\{\frac{2}{r}+\frac{m_0}{r^2}-\frac{2cT_o'}{er}\right\}\right].
\\\nonumber
J_{1(pN)}&=&\frac{r\delta}{c\kappa^2}\left[\left(1+\frac{2m_0}{r}\right)\left\{T_0\left(e-\frac{T_0c}{r}\right)
\left(1-\frac{2m_0}{r}\right)\right\}_{,1}+T_0^2\left(\frac{c}{r}\right)'
\right.\\\nonumber &+&\left.\left(1+\frac{m_0}{r}\right)'
\left(1-\frac{m_0}{r}\right)\left(2T_0e-
\frac{3cT_0^2}{r}+\left(\frac{c}{r}\right)'T_0^2\right)-\frac{2cT_0^2}{r^2}
+\frac{2T_0e}{r}\right.
\\\nonumber&+&
\left.\frac{2cx'T_0'^2}{er}\left(1-\frac{m_0}{r}\right)+
\frac{2cT_0'}{er}\left(2e'-xT_0'\left(1+\frac{m_0}{r}\right)-\frac{4cT_0'}{r}\right)\right].
\end{eqnarray}

Similar to the N approximation strategy, the instability range is
given by
\begin{eqnarray}\label{j15c5}
\Gamma<\frac{\mathcal{A}}{\mathcal{B}},
\end{eqnarray}
where
\begin{eqnarray*}
\mathcal{A}&=&-\frac{\delta T_0
T'_0}{\kappa^2}+\left[\frac{xr}{c}\left(1+\frac{m_0}{r}\right)-\frac{m_0}{r^2}\right]
(\rho_0+p_{r0})-J_{1(pN)}+2(p_{r0}-p_{\perp0})\\\nonumber&\times&\left[\frac{r}{c}
\left(\frac{c}{r}\right)'-\frac{1}{r}\right]-
\frac{m_0}{rc}\left[\frac{c}{r}(3\rho_0+p_{r0}+2p_{\perp0})+J_{0(pN)}\right],\\\nonumber
\mathcal{B}&=&\frac{m_0}{rc}\left[\frac{c}{r}(3\rho_0+p_{r0}+2p_{\perp0})+J_{0(pN)}\right]
+\left[\frac{c}{r}\left(p_{r0}-\frac{\rho_0+p_{r0}}{\rho_0
+p_{\perp0}}p_{\perp0}\right)\right.\\\nonumber&+&\left.\frac{2c}{r}\left(\frac{\rho_0+p_{\perp0}}{\rho_0
+p_{r0}}p_{r0}-p_{\perp0}\right)+
\left(\frac{p_{r0}}{\rho_0+p_{r0}}-
\frac{p_{\perp0}}{\rho_0+p_{\perp0}}\right) J_{0(pN)}\right]
\\\nonumber&+&\frac{r}{c} \left[\frac{c}{r}p_{r0}
+\frac{2c}{r}\frac{\rho_0+p_{\perp0}}{\rho_0
+p_{r0}}p_{r0}+\frac{p_{r0}}{\rho_0+p_{r0}}J_{0(pN)}\right]_{,1}.
\end{eqnarray*}
We obtain instability ranges of collapsing star with vanishing shear
as long as this inequality holds. We analyze that the adiabatic
index evinces the dependence of instability ranges on relativistic
and torsion terms under zero order configuration.

For the dynamical instability condition, it is required that the
right hand side (expressions $\mathcal{A}$ and $\mathcal{B}$) in
inequality (\ref{j15c5}) remains positive. Similarly to the N
regime, we have three cases such as
$\mathcal{A}=\mathcal{B},~\mathcal{A}<\mathcal{B},~
\mathcal{A}>\mathcal{B}$. These cases constitute the instability
ranges as $0<\Gamma<1$ for the first and second case while
$\Gamma>1$ for the last case.

\section{Concluding Remarks}

The collapse in a star occurs due to state of disequilibrium between
inwardly acting gravitational pull and outwardly drawn pressure in
it. In modified theories of gravity, the ranges of dynamical
instability depends on dark source terms in addition to usual terms
(that is terms of GR) determined by the adiabatic index. We have
analyzed the dynamical instability of a collapsing star taking
vanishing shear scalar in ETG gravity. We have taken interior metric
as general spherically symmetric metric while Schwarzschild metric
is considered in exterior region to $\Sigma^{(e)}$ in anisotropic
matter distribution. The contracted Bianchi identities are used to
find two dynamical equations corresponding to a star experiencing
collapse process. A well-known power-law ETG model is considered to
analyze these ranges. We have used perturbation strategy on all
functions such that metric components, energy density, pressure
components, mass, torsion, shear scalar to examine the evolution of
collapse with respect to time. We have applied zero order and first
order perturbed configurations on the field and dynamical equations.
The collapse equations has been constructed through second dynamical
equation. The results are given as follows.

The adiabatic index ($\Gamma$) plays a vital role to examine the
instability ranges for a collapsing star with shear-free condition.
We can analyze the instability regions through this index by
applying N and pN approximations on collapsing equation. In both
cases of approximations, we have observed that this index depends
upon on various quantities such as energy density, anisotropy of
pressure and on some other constraints. This index shows that how we
can modify the instability range of collapsing star under
relativistic as well as torsion terms. Hence, the physical
properties play a crucial role in analyzing the self-gravitating
objects in dynamical instability. Similar to the case of $f(R)$
gravity, the results of present paper contain physical quantities.
It is pointed out that the instability ranges are found as
$0<\Gamma<1$ and $\Gamma>1$ for both approximation i.e., system
yields unstable behavior and meets $\Gamma< \frac{4}{3}$ in the N
and pN perfect fluid limit \cite{S17}.

We would like to point out here that some authors have discussed the
instability analysis by imposing some constraints for finding the
solutions \cite{S19,FS19, S22}-\cite{sr1}. However, we have found
the solutions without imposing any extra constraints.

\end{document}